\documentclass[12pt]{article}

\usepackage{graphicx}
\begin{document}

\smallskip
\centerline{\large \bf Physical behavior of a system representing a particle}

\smallskip
\centerline{\large \bf trapped in a box having flexible size}

\vspace{1.7cm}
\centerline{\bf Yatendra S. Jain}

\smallskip

\centerline{Department of Physics}

\centerline{ North-Eastern Hill University Shillong - 793 022, India}

\vspace{5.0cm}
\bigskip
\begin{abstract}
A critical study of the wave mechanics of a particle in a 1-D box having 
infinite potential walls and small flexibility in its size reveals its  
several important and hither to {\it unknown aspects} which could 
be relevant for a better understanding of systems like {\it quantum} 
-dot/wire/well.  Since most of these aspects arise from the zero point 
force coming into operation when the particle occupies its ground state 
in the box, they are expected to have great significance at low 
temperatures ({\it i.e.}, $T < T_o$, -the $T$ equivalent of the 
ground state energy of the trapped particle).  To demonstrate this 
point, we briefly analyze some important aspects of an electron bubble 
in liquid helium and its nano-droplets which represents a kind of unique 
quantum-dot.  It is argued that our inferences should be equally 
significant for finding a correct microscopic understanding of 
the intriguing behavior of several many body quantum systems such as 
superfluids, superconductors, atomic nucleus, {\it etc.}.

\end{abstract}

\vspace{2.cm}


\bigskip
{\small Email ysjain@email.com}

\vspace{1.5cm}
\copyright by author
\newpage

\noindent
{\bf 1. Introduction}

\bigskip
Soon after an important study of the quantum size effect in semiconductor 
micro-crystals by Ekimov {\it et.al.} [1] reported in 1985, there has been 
manifold development in the field of new systems like {\it quantum}-dot 
/wire/well which assumed importance for their technological 
applications.  While different authors in a recent book [2] elegantly 
review different aspects of these systems, Borovitskaya and Shur [3] 
discuss the basic aspects of the quantum size effects and their origin 
and relate the physical properties of these systems with the wave 
mechanics of a particle ({\it electron}) trapped in a 1-D/2-D/3-D box 
which is analyzed in many elementary texts on wave mechanics [4] and used 
to explain the basics of the above said systems [3].  However, the 
{\it set of standard results} (SSR), [{\it e.g.} Eqns. 1 and 2 
(below) expressing the eigen energy, and eigen function for a particle 
in 1-D box and similar results for 2-D/3-D systems] available in [3,4] 
assume certain {\it ideal situations} : (i) the boundary walls 
of the box are represented by {\it infinite positive potential}, and (ii) the 
dimension(s) (or structure) of the box is considered to have {\it infinite 
rigidity}.  But in the world of {\it real systems}, a trapped particle (say 
an electron in systems like {\it quantum}-dot /wire/well) encounters 
boundary walls of {\it finite height} and the box structure of {\it finite 
rigidity}.  Naturally, the said SSR are not expected to describe such a system 
accurately.  Of course, to a good approximation, one may use the SSR to 
understand the behavior of a system if the particle energy is considerably 
lower than the height of the potential well.  However, since a particle 
trapped in a box exerts a real force (Eqn. 3, below) on the walls of the 
box [4(c), p.67] and produces {\it strain} in the box size ({\it i.e.}, 
{\it finite increase in the size of the box}) and lowers the quantum energies 
of the trapped particle, SSR lose their validity even for low a energy 
particle in a real system which happens to have finite rigidity.  Although,   
the wave mechanics of a particle trapped in a 1-D box of finite rigidity 
has been analyzed in [5] by using WKB approximation, the problem 
has not been studied for several important aspects of the system. For 
example, as concluded here, the strain in the box size not only lowers the 
energy eigen values of the particle but also renders several interesting 
(but {\it hitherto unknown}) properties of the system ({\it cf.} Section 4).  
In addition the physics of a particle in a 1-D is also important for 
understanding the properties of several many body systems, {\it viz.} an 
atomic nucleus [6].  As such we present a critical study of different 
aspects of the system by using two model systems of a particle in a 1-D 
box in Section 2 and summarize our results in Sections 3 and 4.  In near 
future we also plan to complete similar study of a particle trapped in 
2-D and 3-D boxes.  In Section 5 we briefly analyze important experimental 
facts about an electron bubble (a kind of unique quantum-dot) to 
demonstrate the accuracy of our inferences (drawn in Sections 3 and 4)  
and their significance for having a better understanding of systems 
like {\it quantum}-dot /wire/well, helium droplets embedded with a 
particle (electron, atom or a molecule) and other many body quantum 
systems where each constituent particle can be identified with a particle 
trapped in a box.    

\bigskip
\noindent
{\bf 2. Model Systems}

\bigskip
A particle of mass $m$ in a 1-D box of infinite potential walls and size $d$ 
is characterized by its quantum states of energy 
$$E_{\rm n} = \frac{{\rm n}^2h^2}{8md^2}, \eqno(1)$$  

\noindent 
of n-th state (with n = 1,2,3, ... being its quantum number and $h$ the 
Planck constant) represented by the eigenfunction 
$$\Psi_{\rm n} = 
\sqrt{\frac{2}{d}}\sin{({\rm p_nx}/\hbar}),\quad 
{\rm for} \quad 0 < {\rm x} < d, \quad {\rm else}, \quad \Psi_{\rm n} = 0, 
\eqno(2)$$

\noindent
where ${\rm p_n}$ could be identified as the momentum of the particle.  
Two models S1 and S2 of the system (a particle in a 1-D box of infinite 
potential walls) used in the present study are, respectively, shown in 
Fig.1(A) and 1(B). In defining S1 and S2, we identify the following 
facts: 

\smallskip
\noindent
(i) The size of the box in any real system is expected to increase 
under the action of the force [4(c), p.67]
$$F_{\rm n} = -\frac{\partial E_{\rm n}}{\partial d} = 
\frac{{\rm n}^2h^2}{4md^3} \eqno (3)$$ 

\noindent
exerted by the trapped particle on the walls of the box. 
and 

\smallskip
\noindent
(ii) The increase in size $d$ is, obviously, opposed by a force 
$F_{\zeta}$ ({\it Cf.} Eqn 4 below) arising from the 
inherent elasticity of the structure of the box.  

\bigskip
To include (i) and (ii) in S1 and S2, we make an {\it assumption} that 
the left wall of the box is rigidly fixed at x = 0 while the right 
wall located at x =$d$ has some flexibility in its position provided 
and controlled by a spring whose one end is attached to the right 
wall and the other end to a rigidly fixed block.  Since a change in 
the box size means a shift in the position of one wall relative to 
that of the other wall, our assumption remains valid for any 1-D box 
whose two walls may have equal or unequal flexibility for their 
positions under the action of a force $F_{\rm n}$.  When the 
right wall stays at x = $d$, the spring 
has its {\it relaxed state (neither compressed nor extended)}. After 
going through the following discussion, one should find that the 
spring in S1 and S2 merely represents the elasticity of the structure 
of the box which provides the flexibility in its size.  

\bigskip
When $F_{\rm n}$ (Eqn.3) pushes the right wall to its right by $\zeta$ 
(say), the spring gets compressed by equal amount and this calls for a 
force $F_{\zeta}$ by which the compressed spring opposes $F_{\rm n}$ 
and tries to restore the wall at $\zeta = 0$ and we call $F_{\zeta}$ 
as restoring force.  Assuming that $\zeta$ is a small increase 
in the box size, we define $F_{\zeta}$ as 
$$F_{\zeta} = -k\zeta,  \eqno(4)$$  

\noindent
by using Hook's law of elasticity.  While the $-ve$ sign in Eqn.(4) 
signifies that $F_{\zeta}$ tries to decrease $\zeta$,  $k$ represents 
the {\it spring constant} (force per unit $\zeta$).  The origin 
of $F_{\zeta}$ lies with the amount of energy $U$ stored in the 
spring as an increase in its {\it internal energy}.  We have 
$$U = \frac{1}{2}k\zeta^2 = \left[\frac{1}{2}kl^2s^2\right].  \eqno(5)$$

\noindent
where the term in square bracket $[\quad]$ uses $\zeta = sl$ or $s=\zeta/l$ 
which represents the strain in the spring with $l$ being its length 
in its relaxed state.  $U$ is also known as {\it strain energy} of 
the spring for its $s-$dependence (Eqn.5).  For a change in the box 
size from $d$ to $d+\zeta$, $F _{\rm n}$ (Eqn.3) changes to 
$$F _{\rm n}(\zeta) = \frac{n^2h^2}{4m(d+\zeta)^3} \eqno(6)$$

\noindent
which finds equilibrium with $F_{\zeta}$ (Eqn. 4) when 
$\zeta = \delta_{\rm n}$ (say); this implies that the net force on the wall 
$(F_{\rm n}(\zeta) + F_{\zeta})|_{\zeta = \delta_{\rm n}} = 0$ or       
$$\frac{n^2h^2}{4m(d+\delta_{\rm n})^3} - k\delta_{\rm n} = 0, 
\quad {\rm or} \quad \delta_{\rm n} \approx \frac{n^2h^2}{4md^3}\frac{1}{k}
\eqno(7)$$ 

\noindent
where we use $d >> \delta_{\rm n}$ to get $\delta_{\rm n}$.  While 
for $k = \infty$, we find 
$\delta_{\rm n} = 0$ for all n (Case S1, Fig.1(A)) which certifies 
that the box in S1 has a {\it truly fixed} size, for $k \not= \infty$ we 
get $\delta_{\rm n} \not= 0$ (Case S2, Fig.1(B)) indicating that the 
box size changes from $d$ to $d + \delta_{\rm n}$.  This shows that 
S1 is a special case of S2.  To keep the clarity of 
depiction in Fig.1(B), we opt to show the displaced position of the 
right wall only for n = 1 and use $\delta_1 = \Delta d$ to emphasize 
that this much increase in box size $d$ is specifically consequential 
for four important intrinsic aspects of our system(see Sections 4.1-4.4).  

\bigskip
Since a box, whose size assumes no change under the action of 
$F_{\rm n}$, is not expected to exist in nature, S1 can better 
be identified as an {\it ideal system}, while S2 can provide a better 
description to a {\it real system}.   The physical behavior of 
such a system is, therefore, expected to have a better agreement 
with our results ({\it cf.} Sections 3 and 4). However, the 
following analysis implicitly presumes that 
the energy of the particle due to flexibility of the box does not 
change significantly from $E_{\rm n}$ (Eqn.1) and this holds to a 
good approximation only for large $k$ [5] which renders 
$\delta_{\rm n} << d$.  

\bigskip
\noindent
{\bf 3. Common Aspects of S1 and S2}

\bigskip
In a recent study [7] of the wave mechanics of two 
$\delta-$size hard core particles in a 1-D box (also shown to 
be valid for the hard core particles of finite size), we discovered 
that: (i) each particle in a quantum state of the system assumes 
its self-superposition and an effective size of $\lambda/2 = \pi/q$ 
(with $q$ being the magnitude of momentum wave vector of two 
particles having equal and opposite momenta $p$ and $-p$ where 
$p=\hbar q$), (ii) the range of their hard core mutual repulsion 
increases from 
the characteristic 0 value (in classical description) to 
$\lambda/2$ (in quantum description), and (iii) the relative 
motion of two particles in their excited states represents 
collisional motion, while the same in their ground state is found 
to be collision-less.  Since these inferences have an 
important role in finding the {\it intrinsic properties} of our 
system ({\it cf.} Section 4), we rediscover them (Sections 3.2-3.5) 
from a similar analysis of the present system with a view to : (i) affirm 
their relevance to its intrinsic behavior as concluded in Section 4, 
(ii) provide these inferences a stronger foundation, and (iii) have a better 
understanding of the structure of the ground state of an electron 
bubble and the process of its formation ({\it cf.} Section 5).  In 
Section 3.1, we also discuss the signature of wave particle duality 
on the description of $F_{\rm n}$ (Eqn.3) because it helps in 
understanding what we conclude in Sections 3.2-3.5.

\bigskip
\noindent
{\it 3.1 Force} $F_{\rm n}$ {\it and wave-particle duality} 

\bigskip
The fact, that none of the quantity in $F_{\rm n} = -\partial_dE_{\rm n}$ 
(Eqn.3) and  $E_{\rm n}$ (Eqn.1) depends (explicitly or implicitly) on 
time, indicates that $F_{\rm n}$ should be an all time persisting constant 
force that the particle exerts on both the walls of the box.  However, 
we also find the same expression 
$$F_{\rm n} = \frac{2p_{\rm n}}{\tau_{\rm n}} =  
2{\rm p_n}\frac{v_{\rm n}}{2d} = 
\frac{{\rm n}^2h^2}{4md^3}, \eqno (8)$$

\noindent
using a classical model of the particle bouncing back and forth which 
explicitly means that the particle (in its n-th quantum state), moving 
with a velocity $v_{\rm n} = {\rm p_n}/m$, has {\it periodic collision} on 
each wall at time interval ${\tau_{\rm n}} = 2d/v_{\rm n}$ and transfers 
$2{\rm p_n}$ momentum to the wall of the box at its each collision.  
This indicates that $F_{\rm n}$ is a time average of the force(s) exerted 
by the particle on a wall with a periodicity of $\tau_{\rm n}$.  
Alternatively, 
$F_{\rm n}$ is an instantaneous force exerted by the particle during its 
collision and it has an identity with the 
force that a gas molecule (as a classical entity) exerts on a wall of 
its container and contributes to the gas pressure which tends to inflate 
the size of the container.  This means that $F_{\rm n}$ has two 
different descriptions: (i) a time average of forces ({\it as implied by} 
Eqn.8) which agrees with particle nature, and (ii) an all time persisting 
constant force ({\it as evident from the derivation of} Eqn.3) which 
agrees with the wave nature of particle; it may be underlined that only 
due to this nature, a particle in the box assumes discrete quantum states
described by $\Psi_{\rm n}$ (Eqn. 2) and $|\Psi_{\rm n}|^2$ (representing 
the probability of finding the 
particle at a point x) has time independent non-zero values 
at infinitely many different points in the box ($0 \le {\rm x} \le d$).  
We note that the said difference of two 
descriptions of $F_{\rm n}$ can be easily resolved by using 
wave-particle duality, -the particle moves like a wave and leaves 
its impact like a particle which means that our analysis, for the 
first time, demonstrates how a physical quantity like 
$F_{\rm n}$ has two different descriptions due to wave-particle 
duality.  

\bigskip
\noindent
{\it 3.2 Self-superposition of the particle} 

\bigskip
Following the principle of superposition of waves [8], $\Psi_{\rm n}$ 
(Eqn.2) represents the superposition of two plain waves, {\it viz.}, 
$u_{\rm n} = A\exp{(i{\rm p_nx}/\hbar)}$ 
and $w_{\rm n} = A\exp{(-i{\rm p_nx}/\hbar)}$ (with $A$ being the 
normalization constant) of momenta ${\rm p_n}$ and
$-{\rm p_n}$.  We call this superposition as 
{\it self-superposition} of the particle because both these waves  
($u_{\rm n}$ and $w_{\rm n}$) represent {\it one and the same} 
particle.  As discussed 
in [7], particles also have their self-superposition in the 
quantum states of two hard core particles in 1-D box.  However, to 
add further clarity to this meaning, we may mention that the 
self-superposition of a particle, as defined here and in [7], does not 
differ significantly from the {\it self-interference} of a 
particle as defined in [9] because in both cases we have the 
superposition of two plane waves of one and the 
same particle but the two phenomena do differ in their 
physical situations.  In the former case 
$u_{\rm n}$ and $w_{\rm n}$ travel in opposite direction and 
render standing wave like $\Psi_{\rm n}$ (Eqn.2), while in the latter case 
they reach the point of their superposition in nearly the same 
direction. 

\bigskip
\noindent
{\it 3.3 Effective size of the particle}: 

\bigskip
Our experience with classical objects tells us that the size of a 
material particle ({\it say} $b$) means the size of the real space 
it {\it exclusively} occupies.  Evidently the particle of our 
system has $b=0$ since it is implicitly assumed to be a point 
particle.  In wave mechanics, however, a particle is believed to 
manifest itself as a {\it wave packet} [4] of size $\approx \lambda/2$ 
which implies that its $b (\approx \lambda/2)$ has non-zero value 
that depends on its energy/momentum.  Guided by this observation, 
we analyze $E_{\rm n}$ (Eqn.1) and its relation with box size to 
find useful information that improves our understanding of the 
particle size in quantum description. Recasting Eqn.(1), we have
$$E_{\rm n} = \frac{h^2}{8m(d/{\rm n})^2} 
= \frac{s^2h^2}{8m(sd/{\rm n})^2},  \eqno (9)$$

\smallskip
\noindent
which leads us to infer the following:   

\smallskip
\noindent
I(1):  The particle has $E_{\rm n}$ energy also when it occupies 
$s-$th quantum state in the box of size $sd/{\rm n}$ (where the 
integer $s <$ n) indicating that the smallest size of the box 
in which the particle can have $E_{\rm n}$ 
energy is $d/{\rm n}$.

\smallskip
\noindent
I(2): Since $E_{\rm n}$ is equal to the ground state energy of 
the particle for a box of size $d/{\rm n}$, it needs to have a 
higher value, if the box size is reduced below $d/{\rm n}$.  This 
implies that the said particle can not be placed in a box of size 
smaller than $d/{\rm n}$ if its energy $< E_{\rm n}$; in other words, 
the particle with $E_{\rm n}$ energy exclusively occupies a space 
of size $b = d/{\rm n} = \lambda_{\rm n}/2$. 

\bigskip 
Generalizing I(2) we may conclude that a point particle (with 
energy $E$ or momentum p) in its self-superposition state (such as 
$\Psi_{\rm n}$) behaves like 
a wave packet of an {\it effective size} $\lambda/2$ which depends on $E$ 
and p through $\lambda/2 = h/2$p = $h/2\sqrt{2mE}$.  This inference 
not only shows its variance with classical picture where particle size is 
believed to be independent of its $E$ or p but also renders 
$b = \lambda/2$ which differs from $b \approx \lambda/2$ (concluded 
from the definition of wave packet) for its precise magnitude.  
In addition it agrees with an identical inference of our recent 
study on the wave mechanics of two hard core particles in 1-D box [7]. 

\bigskip
\noindent
{\it 3.4 Effective Range of Repulsion}

\bigskip
The particle in our system experiences a potential $V({\rm x})$ 
defined by $V({0 <\rm x} < d)=0$, and $V({\rm x})=\infty$ at all 
other points; in other words it interacts with the walls through 
infinitely strong repulsion only when it presumably occupies x = 0 
or x = $d$ position of the infinite potential wall.  One uses this 
understanding to determine the dynamics of the particle in classical 
(see Ref.[8]) as well as in quantum frameworks [4].  However, 
in variance with the fact that a classical particle in a 1-D square 
box has only kinetic energy, the $d$ dependence of the allowed 
energy values of a quantum particle ($E_{\rm n}$) indicates that 
it can be identified as a potential energy from which we derive 
the force $F_{\rm n}$ (Eqn.3).  In order to examine the nature and  
range of this force we note that the particle can also have same 
$E_{\rm n}$ when the separation between the two walls of the box 
has any of the n-1 possible values ($sd/{\rm n}$ with integer 
$s <$ n) which is valid  for any n (including infinitely large value).    
This indicates that the particle in its n-th state does not 
experience an {\it effective} force which supports or opposes 
a change in the box size in units of $d/{\rm n}$ which has infinitely 
small value for infinitely large n.   

\bigskip
However, when we try to reduce the box size below $d/{\rm n} = 
\lambda_{\rm n}/2$ ({\it i.e.} when the particle is in its 
ground state with energy $E_{\rm n}$), we need to increase 
particle energy above $E_{\rm n}$ since the ground state 
energy of the particle in a box of size $<\lambda_{\rm n}/2$ 
is, obviously, expected to be higher than $E_{\rm n}$.  
This concludes that the particle opposes any reduction in the box 
size below $\lambda_{\rm n}/2$ implying that the 
particle in its ground state experiences a real force 
which pushes it away from the infinite potential walls toward 
$<{\rm x}> = d_{\rm n}/2 = \lambda_{\rm n}/4$ or the particle 
pushes the two walls away from its expected position $<x> = d_{\rm n}/2$.  
Generalizing this inference and using the fact that $<{\rm x}>$ 
in the ground state satisfies $<\rm x> = \lambda/4$, it may 
be concluded that an infinite potential wall experiences the 
pushing action of $F$ (or {\it the particle experiences a repulsion 
from the wall}) when the distance between the particle and the 
wall is $\le \lambda/4$; the potential energy that serves as 
the origin of $F$ is nothing but the ground state energy of a 
particle ($\varepsilon_o$, Eqn.1) which varies as $d^{-2}$.  
Evidently, While $F_1$ always satisfies the condition for its 
persistent pushing action on the two walls of the box, 
$F_{\rm n \ge 2}$ satisfies the condition for its periodic action 
({\it as experienced during periodic collision of particle}).  
Alternatively, the particle experiences persisting impact 
of its interaction ($V({\rm x} = 0/d) = \infty$) with the two 
walls when it rests in its ground state but in the higher 
energy states it experiences such an impact with a periodicity of 
$\tau_{\rm n}$ when its distance from a wall is $\le \lambda/4$. 
    
\bigskip
In summary the range of the impact of the infinite potential walls 
felt by the particle is changed from its {\it zero} value (in 
classical description) to $\lambda/4$ (in quantum description) 
due to wave particle duality.  This agrees with a similar inference 
of our recent study [7] which concludes that two particles, 
interacting though an infinitely strong $\delta-$potential, 
have $<x> \ge \lambda/2$ where $x$ represents the relative 
position of one particle with respect to that of the other.              

\bigskip
\noindent
{\it 3.5 Collisional and collision-less motion}

\bigskip
A classical particle ({\it when made to move}) has collisions 
with walls of the box if its size $b < d$ but such 
collisions cease to exist if $b = d$.  Guided by this observation, 
we use the well known inference of wave mechanics that a particle 
manifests itself as a wave packet of size $\approx \lambda/2$ 
(or to be more precise $\lambda/2$ in the present case, 
{\it cf.}, Section 3.3), we conclude that the particle has: (i) 
periodic collisions (time period $\tau_{\rm n}$) with the walls 
of the box when it occupies higher energy states 
$\Psi_{\rm n \ge 2}$ and (ii) no collisions in $\Psi_1$ because 
{\it the size of its representative wave packet} satisfies 
$\lambda_{\rm n>1}/2 = d/{\rm n} << d$
and $\lambda_1/2 = d$, respectively.  In other words, the 
particle motion ({\it evident from non-zero value of all $E_{\rm n}$ 
representing the expectation values of the kinetic energy operator 
of the particle}) in $\Psi_{\rm  n > 1}$ states could be 
identified as collisional motion, while that in $\Psi_1$ 
as collisionless.  To add clarity to the origin of this 
difference, we note that while $\Psi_{\rm  n > 1}$ having 
n $\ge 2$ anti-nodal loops provide more than {\it one} point (x = 
$(2s+1)\lambda_{\rm n}/4$ with s = 0, 1, 2, 3, n-1) where the 
probability $|\Psi_{\rm  n > 1}|^2$ has maximum and 
identically equal value ($(2/d$) and, as a result of this, the 
particle has a chance to move from one such point to another, 
but the chances of its similar movement do not exist in n=1 
state because $\Psi_1$ has only one anti-nodal loop.  The inference 
is also corroborated by the observation of identical difference 
identified in relation to the forces, $F_1$ and $F_{\rm  n > 1}$, 
exerted by the particle on the walls of the box ($F_1$ acting as an 
all time persisting force, while $F_{\rm n > 1}$ acting periodically 
with a period of $\tau_{\rm n}$, {\it cf.}, Section 3.4). 

\bigskip
\noindent
{\bf 4.  Important Aspects Related to S2}

\bigskip
In this section we try to examine four important aspects of the 
system arising due to the possibility of a change in the box size. 
While three of these ({\it cf.}, Sections 4.1, 4.3 and 4.4) are 
expected from the quantum description (not from classical 
description) of the particle in the box, the fourth related to 
thermal expansion of the box ({\it cf.}, Section 4.2), can be 
expected from classical description but not in the way we find 
from the quantum description.     

\bigskip
\noindent
{\it 4.1  Deviation of Particle energy from} $E_{\rm n}$ 

\bigskip
When the right wall tends to shift from x = $d$ to x = $d + \zeta$ under 
the action of $F_{\rm n}(\zeta)$ against $F_\zeta$, our system to a good 
approximation represents a particle trapped in a 1-D box of impenetrable 
walls.  In the equilibrium state of these forces, the box size becomes 
$d +  \delta_{\rm n}$, since $\zeta = \delta_{\rm n}$ and corresponding 
energy eigenvalues, $E_{S2,\rm n}$, can be obtained by replacing 
$d$ in Eqn.2 with $d+\delta_{\rm n}$ by using the procedure 
followed by Gea-Banacloche for obtaining his {\it Eqns. 6 and 7} in 
[10] for the energies of a particle in a box whose size is approximately 
halved.  To a good approximation this renders
$$E_{S2,{\rm n}} = \frac{n^2h^2}{8md^2(1+\delta_{\rm n}/d)^2}.  \eqno(10)$$ 

\noindent
We note that $E_{S2,{\rm n}} < E_{S1,{\rm n}} \quad ( = E_{\rm n}$, Eqn 1), 
-valid for the ideal case (S1) where the box has infinitely rigid size [4]. 
Subscripts S2 and S1 indicate that the particle energy is related to  
S2 and S1 systems, respectively.  The fall in energy  
$\Delta E_{\rm n}^{(I)} =  E_{S2,{\rm n}} - E_{S1,{\rm n}}$ 
can be obtained, to a first order approximation (indicated by the 
superscript $(I)$), by using $(1+\delta_{\rm n}/d)^{-2} 
\approx 1- 2\delta_{\rm n}/d$ in Eqn.10. We have 

$$\Delta E_{\rm n}^{(I)} = -\frac{4{\rm n}^4E_1^2}{kd^2}. \eqno(11)$$   

\noindent
The fact that it vanishes for $k = \infty$ shows its consistency with 
our inference that S1 is identical to the system studied in [4].  We 
note that similar deviations from the predictions of ideal models 
are also seen in several other cases.  For example, the energy of a 
rotational level of a molecule has lower value than that predicted 
by ideal rigid rotator model [11] because the centrifugal force 
increasing with increasing rotational velocity ({\it i.e.} rotational 
quantum number) increases the molecular dimensions and moments of 
inertia which lower the energy of rotational levels.  

\bigskip
\noindent
{\it 4.2 Strain and Thermal Expansion of the Box}  

\bigskip
In this section we analyze the results of our thought experiment in 
which we monitor the temperature ($T$) dependence of the increase 
in the box size (forced by $F_{\rm n}$, Eqn.3) when our system is 
kept in contact of a thermal bath whose $T$ is slowly reduced to 
zero.  The main objective of this analysis is to demonstrate how 
the said increase in box size and related thermal expansion 
coefficient [$\alpha = (1/d')\partial_Td'$] depends on $T$ 
and at what $T$ the quantum effects dominates the thermal behavior 
of our system.  

\bigskip
Since $F_{\rm n}$ depends on the quantum state occupied by 
the particle which occupies different quantum states at a given $T$ 
with a probability [12] $W_{\rm n} \propto 
Exp{(-E_{\rm n}/k_BT)}$ ($k_B$ being the Boltzmann constant), 
the experimental value of the said increase in the box size should 
be statistical average of $\delta_{\rm n}$ (Eqn.7).  Following 
standard relation for obtaining such an average, it can be 
expressed as    

$$\bar\delta(t)_{QP} = \frac{h^2}{4md^3k}
\frac{\sum_{{\rm n}=1}{\rm n}^2W_{\rm n}}
{\sum_{{\rm n}=1}W_{\rm n}}  \eqno(12)$$

\noindent
where $W_{\rm n}$ representing Maxwell-Boltzmann distribution [13] is 
given by 
$$W_{\rm n} = Ae^{-E_{\rm n}/k_BT} = Ae^{-{\rm n}^2\varepsilon_o/k_BT}
 = Ae^{-{\rm n}^2T_o/T} = Ae^{-{\rm n}^2/t}.  \eqno (13)$$

\noindent
with $A = [\sum_{\rm l=1}^{\infty} \exp{(-E_{\rm l}/k_BT)}]^{-1}$,
(ii) $t=T/T_o$ which represents the temperature of the bath in units 
of $T_o (= \varepsilon_o/k_B)$ (the $T$ equivalent of 
zero point energy $\varepsilon_o$) and (iii) subscript $QP$ to 
emphasize that the particle in the box is a quantum 
particle having discrete $E_{\rm n}$ (Eqn.1) and to 
distinguish it from

$$\bar\delta(t)_{CP} = \frac{2}{kd}
\frac{\int_0^{\infty}EW(E)dE}
{\int_0^{\infty}W(E)dE} = \frac{2\varepsilon_o}{kd}\frac
{\int_0^{\infty}E'e^{-E'/t}dE'}{\int_0^{\infty}e^{-E'/t}dE'} 
\quad {\rm with} \quad E'=\frac{E}{\varepsilon_o}.  \eqno(14)$$  

\noindent
which represents similar quantity for the box containing a {\it classical 
particle}. Eqn.14 uses the fact that the force exerted by a classical 
particle on a wall of 1-D box is $F = 2p/\tau = 2E/(d+\zeta)$ (with $\tau$ 
being the time of a round trip of the box of size $d+\zeta$) which finds its 
equilibrium with the force of spring (= $k\zeta$) at $\zeta = \delta 
\approx 2E/kd$. 

\bigskip
The values of $\bar\delta(t)_{QP}$ (Eqn.12) and $\bar\delta(t)_{CP}$ 
(Eqn.14) calculated in units of $\delta_1 = \Delta d = {h^2}/{4md^3k}$ 
are, respectively, depicted in Fig.2 by Curves A1 and B1 with 
corresponding $\alpha(t)$ depicted by A2 and B2.  While $\bar\delta(t)_{CP}$ 
(Curve B1), having linear dependence on $t$ reaches its zero value at 
$t=0$,  $\bar\delta(t)_{QP}$ (Curve A1), having similar dependence 
for $t > 1$, deviates smoothly at $t \approx 1$ to have unit value 
at $t = 0$.  Consequently, $\alpha(t)$ in the former case remains 
constant for all values of $t$ (Curve B2), while in the latter case it 
deviates around $t \approx 1$ from its constant value at all $t > 1$ 
to assume zero value at $t=0$ (Curve A2). This shows 
that the signatures of the quantum and classical nature of the particle 
on the thermal behavior of the system differ significantly at low 
$T (< T_o)$ at which the particle has very high probability 
[as indicated by $W_{\rm 1} \approx 95\%$ (at $T = T_o$) to 100$\%$ 
(at $T=0$) obtained by using Eq.13 for n=1] to occupy its ground state 
which is characterized by $\lambda/2 = d' \approx d$.     

\bigskip     
It appears that helium atoms in liquid helium also assume a physical state 
which can be identified with a particle trapped in a spherical cavity 
formed by neighboring atoms when the temperature of the liquid 
during the process of its cooling tends to cross $T_o$ at which number 
of particles in their excited states (${\rm n} > 1$) are negligibly 
small. Obviously, under such situations, the $F_1$ is the only force that 
tries to expand the cavity.  To understand the $T-$dependence of the 
strain under these situations with our system, we calculate

$$\bar{\delta}_1(t)_{QP} = 
\frac{h^2}{4md^3k}\frac{W_1}{\sum_{{\rm n}=1}W_{\rm n}}
=\frac{h^2}{4md^3k}\frac{e^{-1/t}}{\sum_{{\rm n}=1}e^{-n^2/t}}  
\eqno(15)$$ 

\noindent
in units of $\delta_1 = \Delta d = {h^2}/{4md^3k}$ and plot it in 
Fig.3 (Curve A) along with the corresponding $\alpha(t)$
(Curve B).  We note that: (i) $\bar{\delta}_1(t)_{QP}$ 
increases smoothly with decreasing $t$ and reaches its maximum 
value $\Delta d$ at $t = 0$ where the particle rests in its ground 
state with $100\%$ probability; in fact as depicted by Curve A, 
$\bar{\delta}_1(t)_{QP}$ assumes $\Delta d$ value closely around 
$t = 1$, and (ii) the corresponding $\alpha(t)$ has $-ve$ values with 
a peak around $t \approx 1$ (Curve B).  

\bigskip
We note that this simple exercise beautifully demonstrates that: 
(i) the $t-$dependence of $\bar{\delta}_1(t)$ and corresponding 
$\alpha(t)$ (Fig. 3) represent unique signature of wave particle 
duality on the thermal expansion of our system when $F_1$ is 
the only operational force, and (ii) $-ve$ 
values of $\alpha(t)$ peaking around $t = 1$ (when observed 
experimentally) should prove that the constituent particle(s) of a 
system represent a particle trapped in a box and the particle in its 
ground state has collision-less motion.  As observed from Fig. 2, the 
results depicted in Fig.3 also reveal that the thermal behavior of 
the system is greatly influenced by the quantum nature of the 
particle when its $T \approx T_o$ or $\lambda/2 \approx d$.  
Interestingly, as discussed in Section 5, these conclusions are 
found to be consistent with the observed $-ve$ thermal expansion of 
liquids $^4He$ and $^3He$ [15].

\bigskip
\noindent
{\it 4.3 Bound state of the particle and the strained box } 

\bigskip
In order to conclude that the particle and the strained box form an 
energetically bound system when the particle occupies the ground state, 
we follow a standard method which can establish whether two atoms 
interacting through certain inter-atomic potential can form a 
diatomic molecule ({\it i.e.} a bound state of two atoms) [16] 
or not.  We start with the total energy of our system where the 
particle rests in its ground 
state in the strained box and we consider that: (i) {\it the particle} 
and (ii) {\it strained box (strained spring}) represent its two 
constituents like two atoms in a diatomic molecule.  
Indicating the ground state by subscript `1', we have    

$$E_{S2,1}(\zeta)  = \frac{h^2}{8m(d + \zeta)^2} +
\frac{1}{2}k\zeta^2 \eqno(16)$$

\noindent
where we use Eqn.(1) for the particle energy in the strained box of 
size $d' = d +\zeta$.  We now determine $\zeta = \Delta d$ (say) for 
which $E_{S2,1}(\zeta)$ has a minimum/ maximum by setting   

$$\frac{\partial E_{S2,1}(\zeta)}{\partial \zeta}|_{\zeta = \Delta d}
 = -\frac{h^2}{4md'^3} + k{\Delta d} =  0 \quad\quad   \eqno(17)$$

\noindent
This renders 

$$k{\Delta d} = 
\frac{2\varepsilon_o}{d}\left[1 + \frac{\Delta d}{d}\right]^{-3}, \eqno(18)$$ 

\noindent
and $E_{S2,1} = \varepsilon_o' + \epsilon_s$ with $\varepsilon_o' = 
h^2/8md'^2$ being the ground state energy of the particle in the 
{\it strained box}, $\epsilon_s = (1/2)k\Delta d^2$ being the 
strain energy of the box, and $d' = d+\Delta d$.  We now determine
$$\frac{\partial^2E_{S2,1}(\zeta)}{\partial \zeta^2}|_{\zeta = \Delta d}
= \frac{3h^2}{4md'^4} + k  \eqno (19)$$

\noindent
whose $+ve$ value establishes that $E_{S2,1}(\zeta)$ has a minimum 
at ${\zeta = \Delta d}$ with a depth which can be found from : (i)
$$\varepsilon_o' = \varepsilon_o\left[1 - 2\frac{\Delta d}{d} + 
3\left(\frac{\Delta d}{d}\right)^2  - 
4\left(\frac{\Delta d}{d}\right)^3 .....\right]   \eqno(20)$$

\noindent
which represents the binomial expansion of 
$\varepsilon_o' = (h^2/8md^2)(1+\Delta d/d)^{-2}$, and (ii)
$$\epsilon_s = \frac{1}{2}k({\Delta d})^2   
 = \varepsilon_o\left[\frac{\Delta d}{d} - 
3\left(\frac{\Delta d}{d}\right)^2 + 
6\left(\frac{\Delta d}{d}\right)^3 - .....\right]
\eqno(21)$$

\noindent
which is obtained by multiplying $\Delta d/2$ with the binomial 
expansion of Eqn.(18).  We find the said depth ($\Delta E_{S2,1}$) 
of minimum in E$_1(\zeta)$ at $\zeta = \Delta d$ by using Eqns.20 
and 21 in 16.  We have $\Delta E_{S2,1} = [E_{S2,1}({\zeta = 
\Delta d}) - E_{S2,1}({\zeta = 0})$] expressed in detail as 
$$\Delta E_{S2,1} = \frac{h^2}{8md'^2} + \frac{1}{2}k{\Delta d}^2 
- \frac{h^2}{8md^2} \approx \varepsilon_o' + \frac{1}{2}k{\Delta d}^2 
- \varepsilon_o  = - \frac{\varepsilon_o\Delta d}{d}   \eqno(22)$$

\noindent
Here we assume that the terms having $\Delta d/d$ with powers more than 2 
in Eqns.20 and 21 are negligibly small.  The fact that $\zeta$ is a 
common factor in: (i) $h^2/8m(d + \zeta)^2$ representing the 
energy of the trapped particle (the kinetic energy of particle 
affected by the presence of infinite potential walls leading to 
$d$ dependence) and (ii) $(1/2)k\zeta^2$ representing 
the strain energy of the box/spring, clearly shows that 
the {\it trapped} particle and the strained box/spring are 
energetically inter-dependent.  Further 
since ${\rm E}({\zeta})$ has a minimum with a depth 
$\Delta E_{S2,1}= - \varepsilon_o\Delta d/d$, the 
two form a physically {\it bound state} like two atoms in a diatomic 
molecule [16].  They remain in this state unless 
$\Delta E_{S2,1}$ energy is supplied from outside.  Although, one 
may similarly find a minimum in $E_{S2,\rm n}(\zeta)$ at $\zeta = 
\delta_{\rm n}$, and corresponding depth $\Delta E_{S2,\rm n} 
= -{\rm n}^2 E_{S1,\rm n}{\delta}_{\rm n}/d$ for any state 
$\Psi_{\rm n > 1}$, but the particle and the strained box/spring would not 
have their stable bound state for n $> 1$ because the particle in 
$\Psi_{\rm n > 1}$ states is free to jump to a lower energy state 
by releasing out the difference in their energies which implies 
that the particle in a $\Psi_{\rm n > 1}$ always has an excess 
energy to overcome corresponding binding energy $-\Delta E_{S2, \rm n}$.  
However, the particle in $\Psi_1$ does not have this option and 
it assumes a stable bound state with the strained box/spring.

\bigskip
\noindent
{\it 4.4 Oscillations and Energy exchange} 

\bigskip
Since, as discussed in Section 4.3, the energies of two constituents 
[{\it (1) trapped particle and (2) the strained box/spring}] of our 
system depend on a 
common factor $\zeta$ and they have their mutually bound state 
for $\zeta = \Delta d$ at which the total sum 
of their energies has minimum value, one expects this system to oscillate 
around $\zeta = \Delta d$ if it is disturbed to have different $\zeta$ 
for a moment.  To study these oscillations, we start our analysis by 
evaluating $E_{S2,1}(\zeta)$ (Eqn.16) for $\zeta = \Delta d$ (the 
point of equilibrium, Eqn.17).  We have 

$$E_{S2,1} =  \frac{h^2}{8md'^2} + \frac{1}{2}k\Delta{d}^2
\eqno(23)$$

\noindent
Assuming that the said equilibrium is disturbed by changing $d'$ to 
$d'\pm \eta$ (with $|\eta| < \Delta{d}$) (Fig 1B), we write 
the corresponding energy as $E_{S2,1}(\eta)$ to have 

$$E_{s2,1}(\eta) =  \frac{h^2}{8m(d'\pm \eta)^2} +
\frac{1}{2}k(\Delta{d \pm \eta})^2  \eqno(24)$$

\noindent
Using 

$$\frac{h^2}{8md'^2}\left[1 \pm \frac{\eta}{d'}\right]^{-2} = 
\frac{h^2}{8md'^2}
\left[1 \mp 2(\frac{\eta}{d'}) + 3 (\frac{\eta}{d'})^2 \mp .......\right] 
\eqno(25)$$

\noindent
in Eqn.24 after dropping all terms containing $\eta/d'$ with powers 
more than two, we find 

$$E_{S2,1}(\eta) \approx \varepsilon_o' +  \epsilon_s
\mp \left[\frac{2\varepsilon_o'}{d'} - k\Delta{d}\right]\eta
+ \frac{1}{2}k'\eta^2         \eqno(26)$$

\noindent
where we use $\epsilon_s = \frac{1}{2}k\Delta{d}^2$ (Eqn.21) and modified 
force constant 

$$k' =  k + \frac{6\varepsilon_o'}{d'^2}  \eqno(27)$$

\noindent
Since the sum $\varepsilon_o' +  \epsilon_s$ (Eqn.27) 
is independent of $\eta$ and the condition for equilibrium (Eqn.17) 
renders  

$$\left[\frac{2\varepsilon_o'}{d'} - k \Delta{d}\right]\eta = 0  \eqno(28)$$
\noindent
we are left with only one $\eta$-dependent term ({\it i.e.}, $k'\eta^2/2$) 
in Eqn.(26).  This concludes that the box (occupied with a particle 
in its ground state) can sustain harmonic oscillations in its size 
and such oscillations 
are governed by an increased value of spring constant $k'$ (Eqn.27). 
Further since $\mp (2\varepsilon_o'\eta/d')$ (Eqn.26) represents a 
linear change (in terms of $\eta$) in $\varepsilon_o'$ and 
$\pm k\Delta{d}\eta$ (Eqn. 26) is a 
similar change in $\epsilon_s = k\Delta d^2/2$ and these changes are 
equal and opposite (Eqn. 28), it is obvious that the particle in 
$\Psi_1$ state and strained box/spring keep 
exchanging energy with each other during $\eta$-oscillations.   

\bigskip
\noindent
{\bf 5. Results and Discussion}

\bigskip
Our results expected to be of great significance to understand 
the quantum behavior of widely different physical systems {\it whose
constituent particles can be identified with a particle trapped in a box}.  
For example, their application to an electron in a quantum-well (a 
thin film of certain semiconductor sandwiched between two slabs of 
other suitable semiconducting material and a good 
representative of a particle ({\it electron}) trapped in a 1-D box) 
can help in having a complete and better understanding of its behavior.  
However, to 
testify our inferences (Sections 3 and 4), we prefer to compare them 
with relevant aspects of an electron bubble [17,18] comprising only 
single electron, although it represents a quantum particle trapped 
in a 3-D spherical cavity.  The extent to which  
our inferences agree with relevant aspects of a quantum-dot /wire/well  
would be discussed in our forth-coming paper since a large number 
of electrons in such a system constitute a gas of 
trapped particles which, obviously, have their collective impact 
on its behavior, -not expected to have a simple 
correlation with our inferences (Sections 3 and 4) for a single 
trapped particle.       

\bigskip
An electron bubble is an electron trapped in a self created cavity 
in helium liquid [17] or its nano-droplets where it is also identified 
as a unique quantum-dot [18].  Under the condition of zero external 
pressure, it is energetically described [19] by  
$$E(R) = \frac{A}{R^2} + BR^2 = \frac{A}{(R_1\pm\zeta)^2} + 
B(R_1\pm\zeta)^2 \eqno(29)$$

\noindent
where $R_1$ represents the radius of the bubble for the electron in 
its ground state identified by subscript $'1'$), $A = h^2/8m_e$ 
(with $m_e$ = mass of electron), and $B = 4\pi\sigma$ (with $\sigma$ 
= surface tension of the liquid). While the first term on the right 
hand side of Eqn.29 is electron energy for its confinement in the 
spherical cavity of radius $R = R_1\pm\zeta$, the second term stands 
for the surface energy of the bubble.  The process of its formation 
with a high energy electron entering into liquid helium (or its nano-droplet) 
and losing its excess energy (above its ground state energy) through 
collisions with surrounding He atoms 
[17-19], indicates that the bubble maintains a spherical shape [17] from 
its incipient state having $R=R_i\approx3.5$ \AA to final state of 
$R=R_1\approx17$\AA.  Since the lowest possible energy 
$E = E_>$ of a particle trapped in a spherical cavity is related to 
cavity radius $R$ through 
$$R = \lambda/2 \quad \quad{\rm with} \quad \quad \lambda = h/\sqrt{2mE_>} 
\eqno (30)$$ 

\noindent 
a decrease in $E_>$ is possible only when $R$ can have corresponding 
increase.  A system like liquid helium provides this possibility 
since a cavity in a liquid is expected to have flexible size because 
its constituent atoms do not have rigidly fixed positions and the 
electron, in accordance with Eqn.(29), exerts a force 
$F(R) = 2AR^{-3}$ on the walls of the cavity and {\it tries to 
expand its size} against the force $F_R = -2BR$ arising from the 
surface energy.  In the state of equilibrium, we have 
$$F(R) + F_R|_{R=R_1} = 0 \quad \quad {\rm or} \quad \quad R_1 = 
\left(\frac{A}{B}\right)^{1/4}   =  
\left(\frac{h^2}{32\pi m_e\sigma}\right)^{1/4}. \eqno (31)$$

\noindent
This shows that the expansion of the bubble from its incipient state 
($R = R_i$ and $E_> = E_i$) to final state ($R = R_1$ and $E_> = E_1$) 
is an act of $F(R)$ which assumes equilibrium with $F_R$ when $R = R_1$ and 
the electron, at every stage of this 
transformation, stays in a state that can be identified as the ground 
state of a particle trapped in a spherical cavity of radius $R$ with 
least possible energy $E_>$ ({\it cf.} Eqn.30) which decreases with 
increase in $R$ unless $R = R_1$.  Since the electron occupies the 
bubble exclusively its effective size ({\it by definition of the size 
of a particle}) can be identified with the size of the bubble which 
means that the electron in a 3-D cavity behaves effectively like a 
spherical body of size ({\it diameter}) $\lambda$ which depends on 
$E_>$ as a result of wave particle duality.  However, on the 
quantitative scale, this size differs by a factor of two from: 
({\it 1}) the effective size ($\lambda/2$) inferred for a particle 
trapped in 1-D cavity ({\it cf.} Section 3.3) as well as ({\it 2}) the 
size ($\approx \lambda/2$) of the representative wave packet of a 
quantum particle.  While it should be interesting to discover the 
reasons for such a difference with wave packet size ({\it 2}), 
particularly when ({\it 1}) matches closely with ({\it 2}), but this 
point would be examined at a later date. 

\bigskip
Use of Eqn.(31) in Eqn.(29) to analyze the state of equilibrium 
({\it i.e.} $R = R_1$ or $\zeta = 0$), renders  
$$E(R_1) = 2\sqrt{AB} = \sqrt{\frac{\pi h^2\sigma}{2m_e}}. \eqno(32)$$

\bigskip
\noindent
which implies that a decrease in $\sigma$ decreases $E(R_1)$, while 
as evident from Eqn.(31), it increases $R_1$ and this agrees with the 
fact that a particle (trapped in a spherical cavity) in its ground 
state satisfies Eqn.(30).  Interestingly, an analysis of 
Eqn.(29) for the excited states, for which $A$ needs to be replaced 
by $A^* = \beta_{n,l}^2A$ (with $n$ and $l$, respectively, 
representing the principal quantum number of the state and angular 
momentum of the particle, and $\beta_{n,l} > 1$ [20]), reveals that 
corresponding $R^*$ should increase with increasing $\beta_{n,l}$ 
indicating that $R^* > R_1$.  Further the fact, that 
excited state energy $E^* > E_1$ and corresponding $\lambda^*/2 < 
\lambda_1/2$, implies that the effective size ($\lambda^*$) of the 
electron in an excited state would be shorter than the bubble 
size $2R_1$ and the electron in such a state can be visualized to 
have collisions with the walls of the bubble cavity even if an 
instantaneous excitation of the electron to such a state does not 
change the $R_1$ to $R^*$.  However, such collisions would not be 
there in the ground state in which effective size of the electron 
$\lambda$ fits exactly with the bubble size $2R$.  We also note that 
the electron in its ground state sits at a distance $R = \lambda/2$ 
from the walls of the cavity which implies that the range of repulsion 
(experienced by the electron with $He$ atoms constituting the walls 
of the cavity) gets increased to $\lambda/2$.    
        
\bigskip
Evidently what follows from the above discussion, the electron bubble 
provides a clear experimental proof for the accuracy of our inferences 
pertaining to: (i) the force that a trapped particle exerts on the walls 
of the cavity/box ({\it cf.} Section 3.1), (ii) the self superposition 
state of a trapped particle ({\it cf.} Section 3.2), (iii) energy 
dependent effective size of a quantum particle ({\it cf.} Section 3.3), 
(ii) the increase in the effective range of repulsion between trapped 
particle and the boundary walls of the cavity ({\it cf.} Section 3.4), 
and (iv) the collisional motion of the particle in its excited state 
and collisionless motion (representing the motion corresponding to 
the zero point energy) in its ground state {\it cf.} Section 3.5).        
 
\bigskip
Eqn.(29) not only represents a situation that can be 
identified with S2 (where structure of the box/cavity has finite 
rigidity, {\it cf.} Section 2) but also matches closely with 
Eqn.16/24 which evidently means that the 
electron bubble has all properties that we concluded in Section 4.  
For example: (i) the {\it structure of the cavity} (an arrangement of 
mutually bound $He$ atoms that constitute the inner layer of the cavity) 
assumes a strain which can be perceived with expanded state of He-He 
bonds in the said structure, (ii) the bubble represents a kind of bound 
state of {\it the trapped electron} and {\it the spherical cavity} 
it creates in helium liquid or its nano-droplets; this is evident 
from the fact that the electron can be separated from the bubble only 
by increasing its energy [18]), (iii) 
the bubble can sustain oscillations ({\it to a good approximation having 
simple harmonic nature}) in its size/radius around $R_1$ and (iv) the 
electron energy for its confinement increases ({\it decreases}) with 
equal loss ({\it gain}) in strain energy of the structure of the cavity 
during such oscillations.  As a strong evidence for the accuracy of 
(ii-iv) we note that the use of $R_1 = (A/B)^{1/4}$ 
(Eqn.31) in Eq.29 renders  
$$E_e(R_1+\zeta) = 2\sqrt{AB} +\frac{1}{2}\left(\frac{6A}{R_1^2}+ 
2B\right)\zeta^2,  \eqno(33)$$ 

\noindent
which has first significant term with only $\zeta^2$ dependence; it 
does not have $\zeta$ dependent term because $E_e(R_1+\zeta)$ has 
minimum value at $\zeta = 0$.  In arranging Eqn.(32) we assume that 
all terms in the binomial expansion of $(A/R_1^2)(1+\zeta/R_1^2)^{-2}$ 
having $\zeta$ with powers $\ge 3$ are negligible.

\bigskip
We may also mention that 3-D systems like liquid $^4He$ and liquid 
$^3He$ exhibit $-ve$ expansion coefficient [15] when their $T$ 
approaches $T_o$ (the $T$ equivalent of the ground state energy of 
a $He$ atom as a particle trapped in a cavity formed by other $He$ 
atoms in the liquid) or the thermal de Broglie wave length [15] a 
$He$ atoms $\lambda_T = h/\sqrt{2\pi mk_BT} (\approx a =$ 
{\it the inter-particle separation}).  Since this condition can be 
compared with $\lambda/2 \approx d$ for the particle in 1-D box, the 
observation of the said $-ve$ expansion provides strong experimental 
support to our inference (Section 4.2) and an added conclusion that the 
particles in these systems at $T \le T_o$ behave like a particle 
trapped in a cavity of size $a$.  

\bigskip
\noindent
{\bf 6.  Conclusions} 

\bigskip
Analyzing the wave mechanics of a particle trapped in a 1-D box 
having small flexibility in its size provided by the elasticity of 
its structure, we discovered its several intrinsic new aspects which 
are expected to help in having a better understanding of the effects 
of wave nature of a particle in several systems like quantum-
(dot/wire/well) [2], electron in an electron bubble [17, 18], 
trapped atom [21], {\it etc.}, as well as the quantum behavior of 
various many body systems.  These aspects are found to have close 
consistency with relevant inferences of our recent study [7].  
Several important aspects of an electron 
bubble provides strong experimental support to our inferences 
(Sections 3 and 4) at qualitative scale; a quantitative agreement is 
not expected because the bubble is a representative of a particle 
trapped in a 3-D (not a 1-D) cavity.  Through the existence of an 
electron bubble, the Nature reveals two important facts : (A1) 
{\it a quantum particle, like an electron in electron bubble,  
exclusively occupies a spherical cavity of radius $R = \lambda/2$ when 
it rests in its lowest possible energy state}, and (A2) {\it such a 
particle exerts a force which 
tries to expand the cavity against the inter-atomic forces among 
the constituents of the cavity wall}.  Although, to the best of our 
knowledge,  A1 and A2 have been experimentally observed so clearly 
for an electron trapped in an electron bubble only but they can not 
be the exclusive properties of only this electron.  In fact these 
should be intrinsic characteristics of any quantum particle.  
For example a conduction electron (in a superconductor) too 
should occupy exclusively a sphere of radius $\lambda/2$, -expected 
to increase with fall in $T$ and in this process it is likely to 
strain the lattice of the conductor by exerting its 
zero-point force against the inter-atomic forces which decide the 
symmetry and structure of the lattice and assume maximum possible 
size when it rests in its ground state in a self created cavity 
in the lattice. In a recent paper [22], we use this possibility to 
reveal the basic foundations of superconductivity and it is 
interesting to note that recent experimental studies [23] do confirm 
the occurrence of lattice strain in superconducting systems.  
In addition, one may find that liquids $^4He$ and $^3He$ exhibit [15] 
$-ve$ volume expansion when they are cooled through $T \approx T_o$ 
which clearly shows that the constituent particles 
($He-$atoms) in their low temperature or superfluid states 
represent a particle trapped in a 3-D cavity.  Guided by 
these facts, we believe that: (i) our important inferences (A1 and 
A2 as stated above) supported strongly by the existence of electron 
bubble would greatly help in having a better understanding of all 
systems like superfluids, superconductors and quantum-dot /wire/well, 
(ii) a microscopic theory of quantum fluidity ({\it superfluidity/ 
superconductivity} [24]) or of similar behavior of other many body 
quantum systems would succeed in providing a complete and correct account 
of these phenomena and related properties if it incorporates (A1) and 
(A2) in its formulation, and (iii) the merit in our theoretical 
work (related to quantum fluidity of a system of interacting 
bosons/fermions [22, 25, 26]) which clearly incorporates (A1) and 
(A2) would have its due recognition sooner or later.  As such we 
make several important and useful inferences ({\it cf.} Sections 3, 4 and 
5) that are strongly supported by experimentally observed existence of an 
electron bubble.  The fact, that the size of the bubble can be identified 
with the effective size of the electron or its representative 
wave packet, rightly emphasizes the importance of wave packet 
and its size in describing a quantum particle in its trapped state and a 
many body quantum system if its constituents behave like a particle 
trapped in cavity formed by neighboring particles.

\newpage

\parindent=-0.75cm
{\bf References}

\vspace{.5cm}

\bigskip
\parindent=-0.75cm 
[1]. A.I. Ekimov, Al. L. Efros, and A.A. Onushchenko, Solid State Commun. 
{\bf 56} (1985), 921-924. 

\bigskip
\parindent=-0.75cm 
[2]. E. Borovitskaya and M.S. Shur, Eds.,  Quantum Dots, 
World Scientific, New Jersey, 2006.

\bigskip
\parindent=-0.75cm 
[3]. Ref. [2], Chapter 1, pp 1-14.

\bigskip
\parindent=-0.75cm 
[4]. For example: (a) L.I. Schiff, Quantum Mechanics, McGraw Hill, Tokyo, 
1968, 3rd ed., and (b) M. S. Rogalski and S.B. Palmer, 
Quantum Physics, Gordon and Breach Science Publications, Australia,1999, 
(c) L.D. Landau and E.M. Lifshitz, Quantum Mechanics 
(Non-relativistic theory),- Course of Theoretical Physics, 
Butterworth Heinemann, Oxford, 1998, Volume 3, 3rd ed., p. 67.  

\bigskip
\parindent=-0.75cm
[5]. K. Tamvakis, Problems and Solutions in Quantum Mechanics, Cambridge 
University Press (2005), pp 80-81.

\bigskip
\parindent=-0.75cm
[6]. V.G. Gueorguiev and J.P. Draayer,  Mixed-Symmetry Shell Model 
Calculations, arXiv:nucl-th/0210034v1 (2002);  V.G. Gueorguiev
Mixed-Symmetry Shell Model Calculations in Nuclear Physics, Ph.D. 
Dissertation, Department of Physics and Astronomy, M.S. Sifia 
University, December 1992.  

\bigskip
\parindent=-0.75cm
[7]. Y.S. Jain, Central European J. Phys. {\bf 2}, 709-719 (2004);a 
small typographical error of this paper has been corrected in its 
e-print available at arxiv.org/quant-ph/0603233 where an added appendix 
also concludes that the expectation value of the hard core interaction 
potential between two particles has zero value for each of their 
quantum states.  

\bigskip
\parindent=-0.75cm
[8]. R.P. Feynman, R.B. Leighton, and M. Sands, The Feynman Lectures on 
Physics, Addison-Wesley Publishing Company, Inc., 1963, Vol. 1, 
Chapters 8-10.

\bigskip
\parindent=-0.75cm
[9]. Ref [8], Chapter. 37.

\bigskip
\parindent=-0.85cm
[10]. J. Gea-banacloche, J. Amer. J. Phys. {\bf 70}, 307 (2002).

\bigskip
\parindent=-0.85cm
[11]. A. Beiser, Perspective of Modern Physics, McGraw-Hill Book 
Company, Auckland, 1969, pp. 316-320.

\bigskip
\parindent=-0.85cm
[12]. D. Halliday, R. Resnick, and J. Walker, Fundamentals of Physics, 
John Wiley and Sons, Inc. New York, 1993, 4th ed., pp 540-541, Eqn 19-10.  

\bigskip
\parindent=-0.85cm
[13]. R. Eisberg and R. Resnick, Quantum Physics of Atoms, Molecules, 
Solids, Nuclei, and Particles, John Wiley and Sons, New York, 2003, 
Eqn. c-6 of Appendix C.

\bigskip
\parindent=-0.85cm
[14]. F. Reiff, Fundamentals of Statistical and Thermal Physics, 
McGraw Hill Book Company, Auckland, 1985, p.12.

\bigskip
\parindent=-0.85cm
[15]. J. Wilks, The Properties of Liquid and Solid 
Helium, Clarendon Press, Oxford, 1967, p. 666 and p. 679. 

\bigskip
\parindent=-0.85cm
[16]. M. Karplus and R.N. Porter, Atoms and Molecules, The 
Benjamin/Cummings Publishing Company, Menlo Park, Chapter 5. 

\bigskip
\parindent=-0.85cm
[17]. M. Rosenblit and J. Jortner, 
Phys. Rev. Lett. {\bf 75}, 4079 (1995)

\bigskip
\parindent=-0.85cm
[18]. M. Farnik, U. Henne, B. Samelin, and 
J.P. Toennies, Phys. Rev. Lett. {\bf 81}, 3892 (1998)

\bigskip
\parindent=-0.85cm
[19]. (i) J. Classen, C.-K. Su, M. Mohazzab and H. Maris, Phys. 
Rev. B {\bf 57}, 3000 (1998), and (ii) H. Maris and S. Balibar, 
Physics Today, {\bf 53} 29 (2000), 

\bigskip
\parindent=-0.85cm 
[20]. S. Flugge, {\it Practical Quantum Mechanics}, Springer-Verlag, 
Berlin (1974), pp 155-159. 

\bigskip
\parindent=-0.85cm
[21]. D.M. Meekhof, C.Monoroe, B.E. King, W.M. Itano and D.J. Wineland,
Phys. Rev. Lett. {\bf 76} (1996), 1796.

\bigskip
\parindent=-0.85cm
[22]. Y.S. Jain, {\it Basic foundations of the microscopic theory of 
superconductivity}, arXiv:Cond-mat/0603784.

\bigskip
\parindent=-0.85cm
[23] A. Bianconi, {\it et.al.}, J. Phys. : Condens. Matter {\bf 12}, 
10655 (2000) and references cited therein;  N.L. Saini, H. Oyanagi, 
A. Lanzara, D. Di Castro, S. Agrestini, A. Bianconi, F. Nakamura and 
T.Fujita, Phys. Rev. Letter. {\bf 64}, 132510 (2001).   

\bigskip
\parindent=-0.85cm
[24]. D.R. Tilley and J. Tilley, {\it Superfluidity and Superconductivity}, 
Adam Hilger Ltd., Bristol (1986).

\bigskip
\parindent=-0.85cm
[25]. Y.S. Jain, Ind. J. Phys. {\bf 79}, 1009 (2005) (Note : The factor 
$|\sin({\bf k}.{\bf r})/2|$ in Eqn. (5) for $\Psi^+$ in this paper should 
be read as $\sin(|{\bf k}.{\bf r}|)/2$ and $E_g(T)/Nk_BT$ in Eqn.(25) 
should be read as $E_g(T)/Nk_B$.

\bigskip
\parindent=-0.85cm
[26]. Y.S. Jain, {\it Macro-orbitals and microscopic theory of a system of 
interacting bosons}, arXiv:Cond-mat/0606571.

\newpage

\bigskip
\begin{figure}
\begin{center}
\includegraphics [angle = 0, width = 0.7\textwidth]{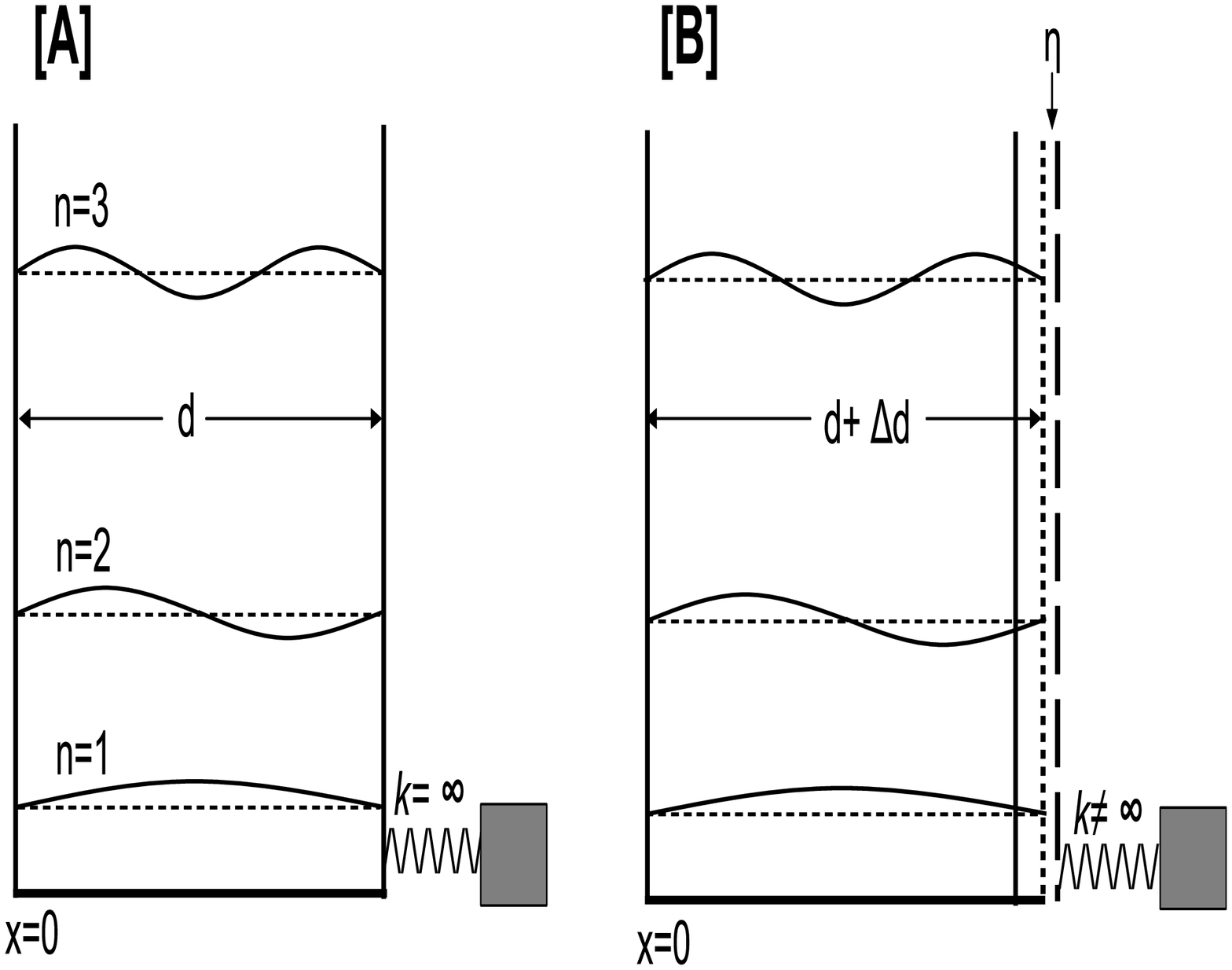}
\end{center}
\end{figure}

\bigskip
\parindent=-1.5cm
Fig. 1: Two model systems of a particle in a box of impenetrable 
walls (see section 2 for details). (A) Model system S1 where both 
walls are rigidly fixed, and (B) model system S2 where left wall 
is rigidly fixed, while the right wall has a flexibility (controlled 
by a spring of finite spring constant $k$) to have some displacement 
to the right.  While $\Delta d$ = displacement of the right wall 
when force $F_1$ (Eqn.4 with n =1) reaches equilibrium with 
$F_{\zeta}$ (Eqn.5), $\eta$ represents an arbitrary but small 
($|\eta| < \Delta d$) displacement from the equilibrium position 
of the right wall from x = $d + \Delta d$ to set oscillation in 
the position of this wall (Section 4.4).

\newpage

\bigskip
\begin{figure}
\begin{center}
\includegraphics [angle = -0, width = 0.7\textwidth]{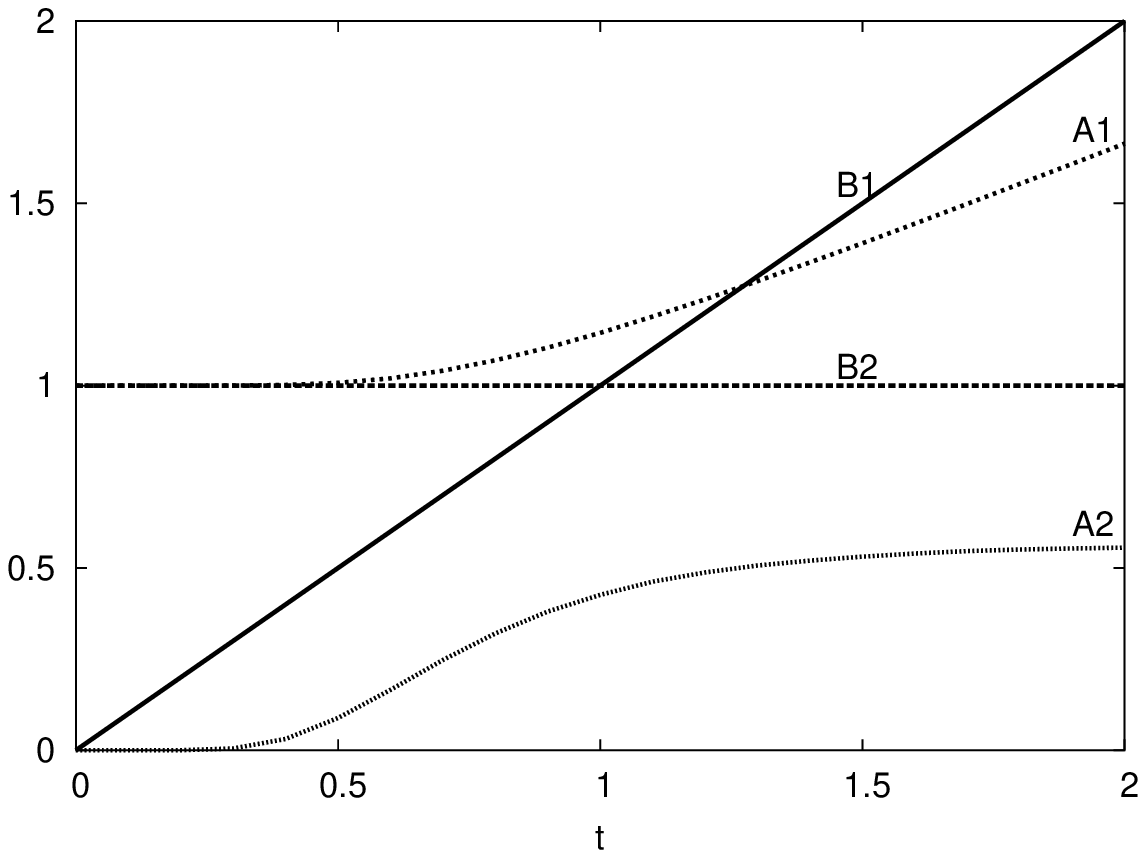}
\end{center}
\end{figure}

\bigskip
\parindent=-1.5cm
Fig. 2: The $t \, (= T/T_o)$ dependence of the increase in the box size. 
While Curve A1 represents $\bar{\delta}(t)_{QP}$ (Eqn.12), Curve B1 
represents $\bar{\delta}(t)_{CP}$ (Eqn.14) obtained in units of 
$\Delta d = \delta_1 = h^2/4md^3k$; corresponding $\alpha(t)$ values are, 
respectively, depicted by Curves A2 and B2 (see Section 
4.2 for details).

\newpage
\bigskip
\begin{figure}
\begin{center}
\includegraphics [angle = -90, width = 0.9\textwidth]{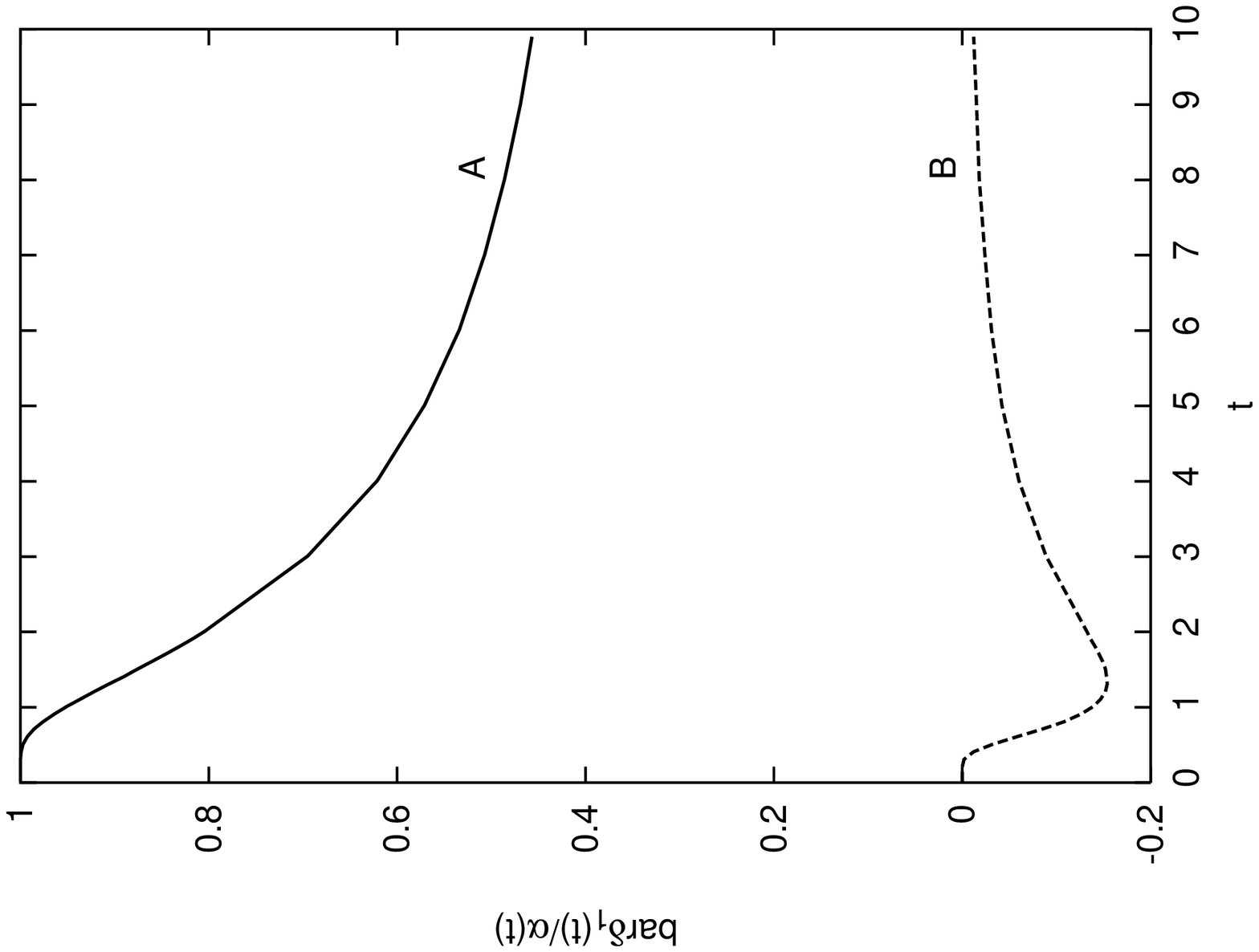}
\end{center}
\end{figure}

\bigskip
\parindent=-1.5cm
Fig. 3: The $t \, (= T/T_o)$ dependence of: (i) the increase in the box size 
$\bar{\delta}_1(t)$ (Curve-A) forced by $F_1$ (Eqn. 4 with n = 1) and 
calculated in units of $\Delta d = \delta_1 = h^2/4md^3k$ by using 
Eqn.15, and (ii) corresponding $\alpha(t)$ (Curve B). 
(See Section 4.2 for details)

\end{document}